\documentclass[reprint,aip,superscriptaddress]{revtex4-1}
\usepackage{amsmath}
\usepackage{amssymb}
\usepackage{cancel}
\usepackage{stmaryrd}
\usepackage{amsfonts}
\usepackage[T1]{fontenc}
\usepackage{bm}
\usepackage{color}
\usepackage{graphicx}
\DeclareSymbolFont{operators}{OT1}{cmr}{m}{n}
\DeclareSymbolFont{letters}{OML}{cmm}{m}{it}
\DeclareSymbolFont{symbols}{OMS}{cmsy}{m}{n}
\DeclareSymbolFont{largesymbols}{OMX}{cmex}{m}{n}
\usepackage{hyperref}
\hypersetup{
    bookmarks=true,         
    unicode=false,          
    pdftoolbar=true,        
    pdfmenubar=true,        
    pdffitwindow=false,     
    pdfstartview={FitH},    
    pdftitle={Metriplectic gyrokinetics},    
    pdfauthor={Eero Hirvijoki},     
    pdfsubject={Collisions in Gyrokinetics},   
    pdfcreator={Eero Hirvijoki},   
    pdfproducer={GNU Make}, 
    pdfkeywords={Collisions} {Gyrokinetics} {Metriplectics}, 
    pdfnewwindow=true,      
    colorlinks=true,        
    linkcolor=blue,      
    citecolor=blue,         
    filecolor=blue,      
    urlcolor=blue           
}
\usepackage{array}
\DeclareFontFamily{OT1}{pzc}{}
\DeclareFontShape{OT1}{pzc}{m}{it}{<-> s * [1.10] pzcmi7t}{}
\DeclareMathAlphabet{\mathpzc}{OT1}{pzc}{m}{it}

\makeatother

\begin{document}

\title{Collisional Gyrokinetics Teases the Existence of Metriplectic Reduction}

\author{Eero Hirvijoki}
\affiliation{Department of Applied Physics, Aalto University, P.O. Box 11100, 00076 AALTO, Finland}
\email{eero.hirvijoki@gmail.com}
\author{Joshua W. Burby}
\affiliation{Los Alamos National Laboratory, Los Alamos, New Mexico 87545, USA}

\date{\today}

\begin{abstract}
In purely non-dissipative systems, Lagrangian and Hamiltonian reduction have proven to be powerful tools for deriving physical models with exact conservation laws. We have discovered a hint that an analogous reduction method exists also for dissipative systems that respect the First and Second Laws of Thermodynamics. In this paper, we show that modern electrostatic gyrokinetics, a reduced plasma turbulence model, exhibits a serendipitous {\it metriplectic} structure. Metriplectic dynamics in general is a well developed formalism for extending the concept of Poisson brackets to dissipative systems. Better yet, our discovery enables an intuitive particle-in-cell discretization of the collision operator that also satisfies the First and Second Laws of thermodynamics. These results suggest that collisional gyrokinetics, and other dissipative physical models that obey the Laws of Thermodynamics, could be obtained using an as-yet undiscovered metriplectic reduction theory and that numerical methods could benefit from such theory significantly. Once uncovered, the theory would generalize Lagrangian and Hamiltonian reduction in a substantial manner.
\end{abstract}

\keywords{Metriplectic dynamics, Gyrokinetics}
 
\maketitle

\section{Inroduction}
Construction of reduced models in physics, when done rigorously, usually results in equations that are expressed in terms of formal infinite series. In order to make practical progress with such models, the series must always be truncated while respecting the essential physics. If the truncation is done carelessly, spurious unphysical effects may appear as, e.g., in the basic formulation of the Burnett equations~\citep{Cercignani_1988_boltzmann}. 

A particularly difficult truncation problem arises in the formulation of collisional gyrokinetics (see, e.g., Refs.~\cite{Brizard_Hahm_2007,Sugama_2017PhPl,Abel_2008PhPl}). Because gyrokinetics is used as a tool for modeling plasma turbulence, it is essential that the effects of collisions in the theory do not lead to a violation of The First and Second Laws of Thermodynamics. Artificial sources or sinks of energy or entropy could drastically alter the steady-state turbulence amplitude, and therefore the predicted turbulence-induced transport levels. On the other hand, due to the complicated dependence of the gyrokinetic equations on the background magnetic field geometry, it is nontrivial to include a collision operator in the theory when starting from first principles~\cite{Brizard_2004PhPl}, and to truncate the expression without violating the First or the Second Law. In contrast, formulating energy and momentum conserving gyrokinetic equations without collisions can now be done systematically using  Lagrangian \cite{Brizard_2000,Sugama_2000,Burby-Brizard:2019PhLA} and Hamiltonian reduction~\cite{Burby_2015PhLA}. 

A resolution to this vexing issue in electrostatic gyrokinetic theory was  put forward in Ref.~\cite{Burby_2015_collisions}, were a collision operator was discovered that addresses the truncation problem without destroying the conservation laws or the H-theorem. The key step in the analysis amounted to being mindful of the basic properties of the Landau collision operator~\cite{Landau:1936}. In other words, a problem-specific trick was found. It would be extremely interesting and useful if there was a more profound theory underlying these results, in particular, a generalization of Lagrangian and Hamiltonian reduction for dissipative systems. If uncovered, such a theory could become an essential tool for unraveling truncation problems in many different areas of physics research \cite{Gorban_2004}. 
 
In this paper, we report on a discovery that suggests a promising candidate for such a general dissipation-compatible truncation tool. Specifically, we show that the modern formulation of collisional electrostatic gyrokinetics exhibits a \emph{metriplectic} structure. This beautiful mathematical framework, discovered amidst the 1980's (see, e.g., Refs.~\cite{Kaufman:1982fl,Kaufman:1984fb,Morrison:1984ca,Morrison:1984wu,Grmela:1984dn,Grmela:1984ea,Grmela:1985jd,Morrison:1986vw}), extends the Poisson bracket formulation of classical mechanics to dissipative systems that obey the First and Second Laws of Thermodynamics. This suggests that collisional gyrokinetics, and other dissipative physical models that obey the Laws of Thermodynamics, may be obtained using an as-yet undiscovered metriplectic reduction theory. Metriplectic reduction, once discovered, would generalize Lagrangian and Hamiltonian reduction in a substantial manner by enabling one to consistently apply, e.g., the perturbation tools common in Lagrangian reduction to dissipative systems while retaining the mathematical stuctures, such as the laws of thermodynamics, intact also after the truncation. 

Further hints towards the existence of such theory have been recently found in the studies of rigid bodies aimed at motion control~\cite{Materassi-Morrison:2018CP} and in studying the general framework of metriplectic dynamics with an application towards dissipative magnetohydrodynamics~\cite{Coquinot-Morrison:2020JPlPh}. Since dissipation-free magnetohydrodanymics and its extensions have been recovered from underlying collisionless two-fluid model using the tools of Lagrangian reduction\cite{Burby-MHD-reduction:2017PhPl}, the discovery of a metriplectic structure for dissipative magnetohydrodynamics is indeed a strong indication of the existence of an underlying metriplectic reduction theory.

Finally, the discovery of a metriplectic structure for the electrostatic gyrokinetic collision operator also admits a particle-in-cell discretization. This discretization is particularly useful as it satisfies the Laws of thermodynamics: the resulting finite-dimensional metric bracket conserves the total energy and dissipates a regularized entropy functional. Furthermore, our proposal for the finite-dimensional collision operator, derived directly from the metric bracket, acts in the 5-D phase-space, includes the finite-Larmor-radius effects, and guarantees ab initio positivity of the numerical distribution function. These are all properties that seemed rather elusive to obtain in a numerical application when the electrostatic gyrokinetic collision operator was first discovered \cite{Burby_2015_collisions} but are now immediately available after indentifying the metriplectic structure. Hence the findings we report are a strong indication that also numerical modeling would greatly benefit from discovering the generic framework of metriplectic reduction.
 
\section{The electrostatic gyrokinetic model}
The model problem we consider is a variant of full-$F$ collisional electrostatic gyrokinetics. The system of equations is
\begin{align}
\label{eq:kinetic-eq}
\frac{\partial F_s}{\partial t}+\{F_s,H^{\text{gy}}_s\}^{\text{gc}}_s&=\sum_{\bar{s}}C^{\text{gy}}_{s\bar{s}}(F_s,F_{\bar{s}}),\\
\label{eq:poisson-eq}
\nabla\cdot\bm{E}&=4\pi(\rho_{\text{\text{gy}}}-\nabla\cdot\bm{P}),
\end{align}
where $F_s$ is the gyroangle-independent gyrocenter distribution function, $\rho_{\text{\text{gy}}}(\bm{x})=\sum_s e_s\int F_s \delta(\bm{X}-\bm{x})\,d\bm{z}^{\text{\text{gc}}}_s$ is the gyrocenter charge density, $\bm{P}$ is the gyrocenter polarization density, and $H^{\text{gy}}_s=K^{\text{\text{gy}}}_s+e_s\varphi$ is the single-gyrocenter Hamiltonian. The single-gyrocenter Poisson bracket $\{\,\cdot\, ,\,\cdot\,\}^{\text{gc}}_s$ of species $s$ is derived by taking an exterior derivative of the symplectic part of the single-gyrocenter Lagrangian one-form, and thus is a genuine Poisson bracket. Explicitly, 
\begin{align}\label{eq:poisson_bracket}
\{F,G\}^{\text{gc}}=&\frac{e}{mc}\left(\frac{\partial F}{\partial \theta}\frac{\partial G}{\partial \mu}-\frac{\partial F}{\partial\mu}\frac{\partial G}{\partial \theta}\right)\nonumber\\
&+\frac{\bm{B}^{*}}{mB_{\parallel}^{*}}\cdot\left(\nabla^{*}F\frac{\partial G}{\partial v_{\parallel}}-\frac{\partial F}{\partial v_{\parallel}}\nabla^{*}G\right)\nonumber\\
&-\frac{c\bm{b}}{eB_{\parallel}^{*}}\cdot\left(\nabla^{*}F\times\nabla^{*}G\right),
\end{align}
with the standard definitions 
\begin{align}
\bm{B}^{*}&=\bm{B}+\frac{mc}{e}\nabla\times\left(v_{\parallel}\bm{b}-\frac{c}{e}\mu\bm{R}\right),\\
\nabla^{*}&=\nabla+\bm{R}\,\partial/\partial\theta,
\end{align}
$\bm{R}$ denoting the Littlejohn's gyrogauge field, and $B_{\parallel}^{*}=\bm{b}\cdot\bm{B}^{*}$.
We remark that the volume element is $d\bm{z}^{\text{\text{gc}}}_s= m_s^{-1}B_{\parallel s}^*\,d^3\bm{X}\,dv_\parallel\,d\mu\,d\theta_s$, 
and therefore implies an integration with respect to the species-$s$ gyrophase.

The function $K^{\text{\text{gy}}}_s$ is the gyrocenter kinetic energy, which may be written entirely in terms of the electric field as
\begin{align}
K^{\text{\text{gy}}}_s=&\frac{1}{2}m v_\parallel^2+\mu |\bm{B}_o|
-e\langle\llbracket\bm{\rho}_o\cdot\bm{E}(\bm{X}+\epsilon\bm{\rho}_o)\rrbracket\rangle\nonumber\\
&-\frac{e^2}{2\mu|\bm{B}_o|}\langle\llbracket \widetilde{\bm{\rho}_o\cdot\bm{E}}(\bm{X}+\epsilon\bm{\rho}_o)\widetilde{\bm{\rho}_o\cdot\bm{E}}(\bm{X}+\bm{\rho}_o)\rrbracket\rangle\nonumber\\
&-\frac{e^2}{2m \omega_c^2}\bm{b}_o\cdot\langle\tilde{\bm{E}}(\bm{X}+\bm{\rho}_o)\times I\tilde{\bm{E}}(\bm{X}+\bm{\rho}_o)\rangle.
\end{align}
Here $\langle\cdot\rangle_{s}=(2\pi)^{-1}\int_0^{2\pi}\cdot\,d\theta_{s}$ denotes the average with respect to the species-$s$ gyroangle, tildes denote the fluctuating part of a gyroangle-dependent quantity, $I=\partial_\theta^{-1}$ is the gyroangle antiderivative,  $\llbracket\cdot\rrbracket=\int_0^1\cdot\,d\epsilon$, and $\bm{\rho}_o$ is the zero'th order (gyroangle-dependent) gyroradius vector. The \emph{net} gyrocenter kinetic energy, which is defined as
\begin{align}
\mathcal{K}(\bm{E})=\sum_s \int K^{\text{\text{gy}}}_s\,F_s\,d\bm{z}_s^{\text{\text{gc}}},
\end{align}
defines the gyrocenter polarization density according to $\bm{P}=-\delta\mathcal{K}/\delta\bm{E}$.

The right-hand-side of the kinetic equation~\eqref{eq:kinetic-eq} is given by the energetically-consistent gyrocenter collision operator~\cite{Burby_2015_collisions}. The expression for $C^{\text{gy}}_{s\bar{s}}(F_s,F_{\bar{s}})$ requires the definitions of the gyrocenter position vectors $\bm{y}_{s}(\bm{z})=\bm{X}+\bm{\rho}_{os}$, the gyrocenter relative velocity vector
\begin{align}\label{eq:relative-velocity}
\bm{w}^{\text{gy}}_{s\bar{s}}&=\{\bm{y}_s,H^{\text{gy}}_s\}^{\text{gc}}_s(\bm{z})-\{\bm{y}_{\bar{s}},H^{\text{gy}}_{\bar{s}}\}^{\text{gc}}_{\bar{s}}(\bm{\bar{z}}),
\end{align}
the scaled projection matrix
\begin{align}\label{eq:scaled-projection}
\mathbb{Q}^{\text{gy}}_{s\bar{s}}(\bm{z},\bm{\bar{z}})&=\frac{\mathbb{P}(\bm{w}^{\text{gy}}_{s\bar{s}}(\bm{z},\bm{\bar{z}}))}{w^{\text{gy}}_{s\bar{s}}(\bm{z},\bm{\bar{z}})}, \qquad \mathbb{P}(\bm{\xi})=\mathbb{I}-\frac{\bm{\xi}\bm{\xi}}{\vert\bm{\xi}\vert^2},
\end{align}
and the three-component collisional flux vector
\begin{align}
\bm{\gamma}^{\text{gy}}_{s\bar{s}}=\int \delta^{\text{gy}}_{s\bar{s}}(\bm{z},\bm{\bar{z}})\mathbb{Q}^{\text{gy}}_{s\bar{s}}(\bm{z},\bm{\bar{z}})\cdot\bm{A}^{\text{gy}}_{s\bar{s}}(\bm{z},\bm{\bar{z}})\,d\bm{\bar{z}}_{\bar{s}}^{\text{gc}},
\end{align}
where the gyrocenter delta function is $\delta^{\text{gy}}_{s\bar{s}}(\bm{z},\bm{\bar{z}})=\delta(\bm{y}_s-\bm{\bar{y}}_{\bar{s}})$, and the vector $\bm{A}^{\text{gy}}_{s\bar{s}}$ is defined according to
\begin{align}\label{eq:A-vector}
\bm{A}^{\text{gy}}_{s\bar{s}}(\bm{z},\bm{\bar{z}})=F_s(\bm{z})\{\bm{\bar{y}}_{\bar{s}},F_{\bar{s}}(\bm{\bar{z}})\}_{\bar{s}}^{\text{gc}}-F_{\bar{s}}(\bar{z})\{\bm{y}_{s},F_{s}(\bm{z})\}_{s}^{\text{gc}}.
\end{align}
With these definitions, the gyroangle averaged collision operator is given as 
\begin{align}\label{eq:collision-operator}
C^{\text{gy}}_{s\bar{s}}(F_s,F_{\bar{s}})=-\frac{\nu_{s\bar{s}}}{2}\left\langle\{y_{s,i},\gamma^{\text{gy}}_{s\bar{s},i}\}_s^{\text{gc}}\right\rangle_{s},
\end{align}
where the symmetric coefficient is $\nu_{s\bar{s}}=4\pi e_s^2e_{\bar{s}}^2\ln\Lambda$. This collision operator conserves total energy and species-wise particle number while producing entropy monotonically. Moreover, when the background field is either axisymmetric or translation symmetric, it conserves the corresponding total momentum. For an explicit proof of the conservation laws and entropy-production property, see Ref.~\cite{Burby_2015_collisions}.

\section{Hamiltonian formulation for collisionless part}
In the absense of collisions, electrostatic gyrokinetic theory naturally has the structure of an infinite-dimensional Hamiltonian system. This structure was first studied by Squire \emph{et. al.} in Ref.~\cite{Squire_2013}. Thus, when the collision integral in~\eqref{eq:kinetic-eq} is dropped, one should expect that the resulting system is Hamiltonian in nature. Because this Hamiltonian structure appears as an essential ingredient in the metriplectic formulation of collisional electrostatic gyrokinetics, we now take the time to summarize it.

As is true of any Hamiltonian system, the Hamiltonian structure of collision-free electrostatic gyrokinetics consists of three parts: (1) the system's infinite-dimensional phase space, (2) the Hamiltonian functional $\mathcal{H}_{\text{GK}}$, and (3) the Poisson bracket $\{\cdot,\cdot\}_{\text{GK}}$. The phase space is the easiest piece. It is not difficult to show that the electrostatic potential may be expressed in terms of moments of the distribution function using the gyrokinetic Poisson equation, i.e. $\varphi=\varphi^*(F)$. Thus, the gyrokinetic Vlasov-Poisson system may be written as a first-order ODE on $F$-space where the time derivative of $F$ is given by the collisionless kinetic equation. It follows that the infinite-dimensional phase space for electrostatic gyrokinetics is just $F$-space. The Hamiltonian $\mathcal{H}$ is slightly less trivial to identify, but may be guessed starting from the energy expressions for kinetic systems with polarization effects~\cite{Morrison_2013_gauge-free-lifting,Burby_2017_finite-dimensional-kinetics}. We have
\begin{align}
\mathcal{H}_{\text{GK}}&=\sum_s \int K^{\text{\text{gy}}}_s F_s\,d\bm{z}^{\text{\text{gc}}}_s+\int \bm{P}\cdot \bm{E}\,d^3\bm{x}+\frac{1}{8\pi}\int |\bm{E}|^2d^3\bm{x}\nonumber\\
&=\sum_s \int H^{\text{\text{gy}}}_s F_s\,d\bm{z}^{\text{\text{gc}}}_s-\frac{1}{8\pi}\int|\bm{E}|^2\,d^3\bm{x}
\end{align}
Note that in this expression the electrostatic potential must be regarded as the unique functional of the distribution function given by solving the gyrokinetic Poisson equation, i.e. $\varphi=\varphi^*(F)$.
Finally, the following expression gives the Poisson bracket of two functionals $\mathcal{F}(F)$ and $\mathcal{G}(F)$:
\begin{align}\label{anti_symm_bracket}
\{\mathcal{F},\mathcal{G}\}_{\text{GK}}=\sum_s \int \left\{\frac{\delta\mathcal{F}}{\delta F_s},\frac{\delta\mathcal{G}}{\delta F_s}\right\}^{\text{\text{gc}}}_s\,F_s\,d\bm{z}^{\text{\text{gc}}}_s,
\end{align}
which represents a convenient simplification of the first reported bracket~\cite{Squire_2013}. 
In fact, Eq.\,\eqref{anti_symm_bracket} is an example of a so-called Lie-Poisson bracket~\cite{Marsden_Ratiu_2010}, which is perhaps the simplest non-canonical bracket one would expect to see in a continuum field theory. Here the functional derivative of an observable $\mathcal{A}(F)$ is the unique gyroangle-independent function of $(\bm{X},v_\parallel,\mu)$ such that
\begin{align}
\delta\mathcal{A}(F)=\sum_s\int \frac{\delta\mathcal{A}}{\delta F_s}\,\delta F_s\,d\bm{z}^{\text{\text{gc}}}_s
\end{align}
for arbitrary variations $\delta F_s$.

This Hamiltonian structure is related to collisionless gyrokinetic dynamics as follows. Given a functional $\mathcal{Q}$ on $F$-space, the dynamics of $\mathcal{Q}(F)$, with $F$ evolving according to the electrostatic gyrokinetic Vlasov-Poisson system, are specified by
\begin{align}\label{Hamilton_eq}
\frac{d\mathcal{Q}}{dt}=\{\mathcal{Q},\mathcal{H}_{\text{GK}}\}_{\text{GK}}.
\end{align}
By choosing $\mathcal{Q}(F)=\int \delta(\bm{z}-\bar{\bm{z}})\,F(\bar{\bm{z}})\,d\bar{\bm{z}}^{\text{gc}}$, Eq.~\eqref{Hamilton_eq} reproduces the collisionless limit of~\eqref{eq:kinetic-eq}. The least straightforward step in demonstrating this fact is showing that the functional derivative of $\mathcal{H}_{\text{GK}}$ is the single-gyrocenter Hamiltonian $H^{\text{\text{gy}}}_s$, i.e. $H^{\text{\text{gy}}}_s=\delta\mathcal{H}_{\text{GK}}/\delta F_s$. To see that this is so, observe that an arbitrary variation of the gyrokinetic system Hamiltonian is given by
\begin{align}
\delta\mathcal{H}_{\text{GK}}=&\sum_s \int H^{\text{\text{gy}}}_s\delta F_s\,d\bm{z}^{\text{\text{gy}}}_s \nonumber\\
&+\int(\rho_{\text{\text{gy}}}-\nabla\cdot\bm{P}-(4\pi)^{-1}\nabla\cdot\bm{E})\,\delta\varphi\,d^3\bm{x},
\end{align}
where the variation of the electrostatic potential $\delta\varphi$ is a complicated linear functional of $\delta F$. Because $\varphi=\varphi^*(F)$ in the gyrokinetic Hamiltonian, the gyrokinetic Poisson equation may now be used to kill the second term, and thereby deduce the desired result.

\section{Metric bracket for the collision operator}
Metriplectic dynamics \cite{Kaufman:1982fl,Kaufman:1984fb,Morrison:1984ca,Morrison:1984wu,Grmela:1984dn,Grmela:1984ea,Grmela:1985jd,Morrison:1986vw} provides a convenient framework to describe systems that exhibit both Hamiltonian and dissipative character. The Hamiltonian contribution in such a system is represented in terms of an energy functional ${\cal H}$ and an antisymmetric Poisson bracket $\{\, \cdot \, , \, \cdot \, \}$ while the dissipative contribution is represented in terms of an entropy functional ${\cal S}$ and a symmetric, negative semi-definite metric bracket $(\, \cdot \, , \, \cdot \, )$. When combined, the evolution of a given functional ${\cal Q}$ is obtained from the equation
\begin{align}
\frac{d {\cal Q}}{d t}=\{{\cal Q},{\cal F}\}+({\cal Q},{\cal F}),
\end{align}
where ${\cal F}={\cal H}-{\cal S}$ denotes a generalized free-energy functional that is dissipated via increase in the system entropy. 

For this framework to respect the laws of thermodynamics, one requires ${\cal H}$ to be an invariant of the metric bracket and ${\cal S}$ an invariant of the Poisson bracket in the sense of $({\cal H},{\cal A})=0$ and $\{{\cal S},{\cal A}\}=0$ with respect to an arbitrary functional ${\cal A}$. Furthermore, ${\cal S}$ must not be an invariant of the metric bracket. Then, it is straight forward to verify the properties $d{\cal F}/dt\le 0$, $d{\cal H}/dt=0$, and $d{\cal S}/dt\ge 0$. The system may display also other invariants $\{{\cal C}_i\}_i$ which are invariants of the total bracket. In an equilibrium state, $d{\cal Q}/dt=0$ for all possible ${\cal Q}$. One way for such a state to exist is that the free-energy functional is a linear combination of the {\it common} invariants of the two brackets according to
\begin{align}
{\cal F}=\sum_i c_i\,{\cal C}_i,
\end{align}
where the coefficients $c_i$ are uniquely determined by the initial state of the system. 

To find a corresponding metriplectic formulation for electrostatic gyrokinetics, we have to dress the collision operator in terms of a symmetric bracket. Fortunately, this turns out to be a rather straight forward task, once we employ the identity $\delta {\cal H}^{\text{gy}}/\delta F_s=H^{\text{gy}}_s$. 

In the expression \eqref{eq:A-vector} for the vector $\bm{A}_{s\bar{s}}^{\text{gy}}(\bm{z},\bm{\bar{z}})$, one may identify functional derivatives $\delta {\cal S}_{\textrm{GK}}/\delta F_s$ of an entropy functional
\begin{align}
{\cal S}_{\textrm{GK}}=-\sum_s \int F_s(\bm{z})\ln F_s(\bm{z}) d\bm{z}^{\text{gc}}_s.
\end{align}
A weak form of the collision operator \eqref{eq:collision-operator}, and some further reasoning, then summon a symmetric, negative semi-definite bracket
\begin{align}
({\cal A},{\cal B})_{\textrm{GK}}=-\sum_{s\bar{s}}\frac{\nu_{s\bar{s}}}{4}\iint\bm{\Gamma}^{\text{gy}}_{s\bar{s}}({\cal A})\cdot\mathbb{W}^{\text{gy}}_{s\bar{s}}\cdot\bm{\Gamma}^{\text{gy}}_{s\bar{s}}({\cal B})d\bm{z}^{\text{gc}}_{\bar{s}}d\bm{z}^{\text{gc}}_s,
\end{align}
where the vector $\bm{\Gamma}^{\text{gy}}_{s\bar{s}}({\cal A})$ and the symmetric, positive semi-definite tensor $\mathbb{W}_{s\bar{s}}^{\text{gy}}$ are
\begin{align}
\bm{\Gamma}^{\text{gy}}_{s\bar{s}}({\cal A})&=\left\{\bm{y}_{\bar{s}},\frac{\delta {\cal A}}{\delta F_{\bar{s}}}\right\}^{\text{gc}}_{\bar{s}}(\bm{\bar{z}})-\left\{\bm{y}_s,\frac{\delta {\cal A}}{\delta F_s}\right\}^{\text{gc}}_{s}(\bm{z}),\\
\mathbb{W}_{s\bar{s}}^{\text{gy}}&=\delta^{\text{gy}}_{s\bar{s}}(\bm{z},\bm{\bar{z}})\mathbb{Q}_{s\bar{s}}(\bm{z},\bm{\bar{z}})F_s(\bm{z})F_{\bar{s}}(\bm{\bar{z}}).
\end{align}
It is straight forward to verify that evaluation of the bracket $({\cal Q}_s,-{\cal S}_{\textrm{GK}})_{\textrm{GK}}$, with respect to ${\cal Q}_s=\int \delta(\bm{\tilde{z}}-\bm{z})F_s(\bm{z})\,d\bm{z}^{\text{gc}}_s$ leads to the expression \eqref{eq:collision-operator} evaluated at $\bm{\tilde{z}}$. 

To complete a metriplectic formulation for the electrostatic gyrokinetic model, we need to verify that the electrostatic energy functional ${\cal H}_{\textrm{GK}}$ is an invariant of the metric bracket and that the entropy functional ${\cal S}_{\textrm{GK}}$ is not.
This is straight forward to demonstrate: Since $\delta {\cal H}_{\textrm{GK}}/\delta F_s = H^{\text{gy}}_s$, we have
\begin{align}
({\cal H}_{\textrm{GK}},{\cal B})_{\textrm{GK}}&=\sum_{s\bar{s}}\frac{\nu_{s\bar{s}}}{4}\iint 
\bm{w}^{\text{gy}}_{s\bar{s}}\cdot\mathbb{W}^{\text{gy}}_{s\bar{s}}\cdot\bm{\Gamma}^{\text{gy}}_{s\bar{s}}({\cal B})d\bm{z}^{\text{gc}}_s d\bm{\bar{z}}^{\text{gc}}_{\bar{s}}\nonumber\\
&=0,
\end{align}
where ${\cal B}$ is an arbitrary functional. This follows from the property $\bm{w}^{\text{gy}}_{s\bar{s}}\cdot\mathbb{W}^{\text{gy}}_{s\bar{s}}=0$. Entropy, on the other hand, is not an invariant of the metric bracket since a correct choice for ${\cal B}$ reproduces the expression for the collision operator as stated above. We may thus conclude that the system consisting of equations~\eqref{eq:kinetic-eq} and~\eqref{eq:poisson-eq} exhibits a metriplectic structure, and that the dynamics of any functional is given by
\begin{align}
\frac{d{\cal Q}}{dt}=\{{\cal Q},{\cal F}_{\textrm{GK}}\}_{\textrm{GK}}+({\cal Q},{\cal F}_{\textrm{GK}})_{\textrm{GK}},
\end{align}
where ${\cal F}_{\textrm{GK}}={\cal H}_{\textrm{GK}}-{\cal S}_{\textrm{GK}}$ is the gyrokinetic free-energy functional.

We note that if the background magnetic field is axially symmetric, the total toroidal angular momentum
\begin{align}
{\cal P}_{\phi}=\sum_s\int p_{\phi s}(\bm{z})F_s(\bm{z}) d\bm{z}^{\text{gc}}_s,
\end{align}
is an invariant of the metric bracket, though not of the Poisson bracket. This can be verified as follows: Since $\delta {\cal P}_{\phi}/\delta F_s=p_{\phi s}$, with $p_{\phi s}$ the single-particle guiding-center canonical momentum of species $s$, the expression
\begin{align}
&({\cal P}_{\phi},{\cal B})_{\textrm{GK}}=\sum_{s\bar{s}}\frac{\nu_{s\bar{s}}}{4}\iint\bm{\Gamma}^{\text{gy}}_{s\bar{s}}({\cal P}_{\phi})\cdot\mathbb{W}^{\text{gy}}_{s\bar{s}}\cdot \bm{\Gamma}^{\text{gy}}_{s\bar{s}}({\cal B})d\bm{z}^{\text{gc}}_s d\bm{\bar{z}}^{\text{gc}}_{\bar{s}}\nonumber\\
&=\sum_{s\bar{s}}\frac{c_{s\bar{s}}}{4}\iint\bm{e}_{z}\times\left(\bm{\bar{y}}_{\bar{s}}-\bm{y}_{s}\right)\cdot \mathbb{W}^{\text{gy}}_{s\bar{s}}\cdot \bm{\Gamma}^{\text{gy}}_{s\bar{s}}({\cal B})d\bm{z}^{\text{gc}}_s d\bm{\bar{z}}^{\text{gc}}_{\bar{s}}\nonumber\\
&=0
\end{align}
vanishes identically since the integrand contains the term $\left(\bm{\bar{y}}_{\bar{s}}-\bm{y}_{s}\right)\delta^{\text{gy}}_{s\bar{s}}(\bm{z},\bm{\bar{z}})$.

Thermodynamical equilibrium is reached once the differential of the free-energy functional is a linear combination of the invariants of the total metriplectic bracket evaluated at the equilibrium. The simplest way to achieve this condition is to have the free-energy differential be an invariant of the Poisson bracket and the metric bracket individually, which happens to imply the differential of $\mathcal{F}$ is the differential of the sum of the total mass of each species. From this condition, the equilibrium distributions $F_{\text{eq},s}$ may be solved by taking variations, which leads to
\begin{align}
F_{\text{eq},s}\sim\exp\left(-\frac{H^{\text{\text{gy}}}_{s}}{T}\right),
\end{align}
with common temperature for each species.

\section{Particle-in-cell discretization for the collisions}
As a final note, we discuss how the newly discovered bracket could be discretized with marker particles to provide a meaningful finite-dimensional approximation of the collision operator that acts on individual particle's phase-space position in a deterministic manner. To avoid extra clutter of indices, we now consider only the single-species case while it is straightforward to extend the following results to the multiple species as well.  

Say we have chosen to represent the phase-space density with marker particles, so that the distribution function multiplied by the phase-space Jacobian is parametrized by 
\begin{align}\label{eq:finite-distribution}
    m^{-1}B_{\parallel}^{\ast}(\bm{z}) F(t,\bm{z})=\sum_{p}n_p\delta(\bm{z}-\bm{z}_p(t)).
\end{align}
In a particle-in-cell approach, the Hamiltonian motion would push the locations $\bm{z}_p(t)$ forward in time along the characteristics generated by the vector  $\dot{\bm{z}}_p=\{\bm{z},H^{\text{gy}}\}^{\text{gy}}(\bm{z}_p)$ while $n_p$---the number of real particles each marker carries or, more commonly, the particle weight---are arbitrary constants sampled from the initial phase-space density. To find an analogous vector accounting for the effect of collisions in individual marker-particle motion, our finite-dimensional metric bracket of any two functions $A$ and $B$ that depend on $n_p$ and $\bm{z}_p$ will read
\begin{align}\label{eq:finite-bracket}
    (A,B)=-\frac{\nu}{4}\sum_{p,\bar{p}}\left\langle\langle\bm{\Gamma}(A;p,\bar{p})\cdot \mathbb{W}_{\epsilon}(p,\bar{p})\cdot\bm{\Gamma}(B;p,\bar{p})\right\rangle\rangle_{p,\bar{p}}.
\end{align}
Here $\langle\langle\,\cdot\,\rangle\rangle_{p,\bar{p}}$ refers to double gyroaverage over both the gyroangles of particles $p$ and $\bar{p}$ which can be performed as a quadrature or an $n$-point average (implementing a quadrature or the average will not affect the conservation laws nor the entropy dissipation). The vector $\bm{\Gamma}(A;p,\bar{p})$ in \eqref{eq:finite-bracket} is defined according to 
\begin{align}\label{eq:finite-gamma}
    \bm{\Gamma}(A;p,\bar{p})=\frac{\{\bm{y},A\}^{\text{gc}}(\bm{z}_{p})}{n_p}-\frac{\{\bm{y},A\}^{\text{gc}}(\bm{z}_{\bar{p}})}{n_{\bar{p}}},
\end{align}
and the single-particle Poisson bracket is now understood in the sense of
\begin{align}
    \{\bm{y},A\}^{\text{gc}}(\bm{z}_{p})=\{\bm{y},z^\alpha\}^{\text{gc}}(\bm{z}_p)\frac{\partial A(...,\bm{z}_p,...)}{\partial z^{\alpha}_p}.
\end{align}
Hence, if $A$ is independent of some specific $\bm{z}_k$, then $\{\bm{y},A\}^{\text{gc}}(\bm{z}_{k})=0$. 
The matrix $\mathbb{W}(p,\bar{p})$ in \eqref{eq:finite-bracket} is given by
\begin{align}
    \mathbb{W}_{\epsilon}(p,\bar{p})=n_{p}n_{\bar{p}}\delta_{\epsilon}(p,\bar{p})\mathbb{Q}(\bm{z}_p,\bm{z}_{\bar{p}}),
\end{align}
where $\mathbb{Q}(\bm{z}_p,\bm{z}_{\bar{p}})$ is the scaled projection matrix constructed from the single-particle Hamiltonian the same way as previously. The major difference to the infinite-dimensional bracket is the approximation of the strict delta function $\delta^{\text{gy}}(\bm{z}_p,\bm{z}_{\bar{p}})$ with, e.g., a parametrized radial basis function $\delta_{\epsilon}(p,\bar{p})=\Psi_{\epsilon}(\bm{y}(\bm{z}_p)-\bm{y}(\bm{z}_{\bar{p}}))$. This approximation is mandatory to account for the finite difference in the particle locations. In a numerical application, it is used to spatially screen which of the particles collide with each other, effectively representing the size of a spatial collision "cell".

The finite-dimensional total energy, now constructed as the weighted sum $H=\sum_pn_pH^{\text{gy}}(\bm{z}_p(t),t)$ and the explicit time dependence referring to the electrostatic potential, is trivially an invariant of the finite-dimensional bracket \eqref{eq:finite-bracket} for $\bm{\Gamma}(H;p,\bar{p})$ is the null eigenvector of the matrix $\mathbb{W}_\epsilon(p,\bar{p})$. Unfortunately the finite-dimensional expression for the Canonical angular momentum $P_{\phi}=\sum_p n_p p_{\phi}(\bm{z}_p(t))$ is not an exact invariant. The approximation of the strict delta function $\delta^{\text{gy}}(\bm{z}_p,\bm{z}_{\bar{p}})$ with the radial basis function $\delta_{\epsilon}(p,\bar{p})$, introduced to account for the localization of the particles' positions, destroys this property, though in a controlled manner with the error being quantified by the chosen width of the function $\delta_{\epsilon}(p,\bar{p})$. 

Finally, to obtain dynamics, an approximative entropy functional is required. This can be done by, e.g., convoluting the delta distribution \eqref{eq:finite-distribution} with some phase-space radial basis function $\Phi_{\epsilon}(\bm{z})$ so that the finite-dimensional entropy is expressed as
\begin{align}
    S&=\int \frac{\sum_p n_p\Phi_{\epsilon}(\bm{z}-\bm{z}_p)}{m^{-1}B_\parallel^\ast(\bm{z})}
    \ln\left[\frac{\sum_{\bar{p}}n_{\bar{p}}\Phi_{\epsilon}(\bm{z}-\bm{z}_{\bar{p}})}{m^{-1}B_{\parallel}^{\ast}(\bm{z})}\right]d\bm{z}^{\text{gc}}.
\end{align}

The finite-dimensional bracket \eqref{eq:finite-bracket} will then always dissipate the approximate entropy functional for the bracket is negative semidefinite. While an arbitrary choice of $\Phi_{\epsilon}$ will likely not provide a numerical H-theorem, specifically the equilibrium state, a proper choice of the convolution function $\Phi_{\epsilon}$ might succeed in the feat the same way as was recently demonstrated for the particle Landau operator \cite{Carrillo_et_al:2020} using the Gaussian radial basis function and its special properties under convolution. With respect to 
the chosen, regulated entropy functional, the collisional particle characteristics are then obtained from $(\dot{A})_{\text{collisions}}=(A,S)$ by substituting $A=\bm{z}_p$. The result, upon using the antisymmetry of $\bm{\Gamma}$ and the symmetry of $\mathbb{W}$, can be written as
\begin{align}\label{eq:collisional-motion}
    \dot{\bm{z}}_p=-\frac{\nu}{2}\left\langle\{y^i,\bm{z}_p\}^{\text{gc}}U^i(p)\right\rangle_p,
\end{align}
where the vector $\bm{U}(p)$ is defined according to
\begin{align}\label{eq:u-vector}
    \bm{U}(p)=\sum_{\bar{p}}n_{\bar{p}}\left\langle\delta_{\epsilon}(p,\bar{p})\mathbb{Q}(\bm{z}_p,\bm{z}_{\bar{p}})\cdot\bm{\Gamma}(S,p,\bar{p})\right\rangle_{\bar{p}}.
\end{align}

To evaluate the collisional rate-of-change of the particle coordinates, it is thus necessary only to evaluate the expressions $\{X^i+\rho_o^i,z^{\alpha}\}^{\text{gc}}$, $\partial_{\alpha}H^{\text{gy}}$, and $\partial_{\alpha}S$ for each particle and to combine them to the expressions $\mathbb{Q}^{ij}(\bm{z}_p,\bm{z}_{\bar{p}})$ and $\Gamma^i(S,p,\bar{p})$, and ultimately to the vector $U^i(p)$. Out of the necessary expressions, $\{X^i,z^\alpha\}$ and $\partial_\alpha H^{\text{gy}}$ are needed also in integrating the Hamiltonian trajectories for $\{X^i,z^{\alpha}\}^{\text{gc}}\partial_\alpha H^{\text{gy}}\equiv \dot{X}^i$ represents the gyrocenter's parallel and drift velocities while the extra term $\{\rho_o^i,z^{\alpha}\}^{\text{gc}}\partial_\alpha H^{\text{gy}}\equiv\dot{\rho}^i_o$ is effectively the part of particle's perpendicular velocity vector resulting from the local cyclotron motion but expressed in terms of the gyrocenter coordinates. The double gyroaverage can be computed by sampling, say, four values for each particle's gyroangle, or even by just one value. The energy-conservation property is not affected by this choice for the exploit of the nullspace of $\mathbb{Q}$ happens point-wise inside the double gyroaverage.  

For convenience, we list here the necessary Poisson-bracket expressions in Cartesian coordinates. Using \eqref{eq:poisson_bracket}, they read 
\begin{align}
    \{X_i+\rho_{o,i},X_j\}^{\text{gc}}&=\varepsilon_{j\ell k}\frac{cb_k}{eB_{\parallel}^\ast}(\delta_{\ell i}+\partial^\ast_\ell\rho_{o,i}),
    \\
    \{X_i+\rho_{o,i},v_\parallel\}^{\text{gc}}&=\frac{B^\ast_k}{m B_{\parallel}^\ast}\left(\delta_{ki}+\partial_k^\ast\rho_{o,i}\right),
    \\
    \{X_i+\rho_{o,i},\mu\}^{\text{gc}}&=\varepsilon_{ijk}\rho_{o,j}\frac{eb_k}{mc},
\end{align}
where the components of the dyad $\nabla^\ast\bm{\rho}_o$ are 
\begin{align}
    \partial_i^\ast\rho_{o,j}=-\frac{1}{2}\partial_i\ln B\,\rho_{o,j}-\partial_ib_k\,\rho_{o,k}b_j.
\end{align}
and $\varepsilon_{ijk}$ is the Levi-Civita tensor. Given the the derivatives of the Hamiltonian, $\partial_\alpha H^{\text{gy}}$, and of the entropy functional, $\partial_\alpha S$, with respect to the coordinates $z^\alpha_p$ and $z^\alpha_{\bar{p}}$ of two particles $p$ and $\bar{p}$, it is then a straightforward task to use the given Poisson-bracket expressions and the generic property $\{f,g\}=\{f,z^\alpha\}\partial_\alpha g$ to construct the relative velocity vector \eqref{eq:relative-velocity} needed for the matrix $\mathbb{Q}(\bm{z}_p,\bm{z}_{\bar{p}})$ \eqref{eq:scaled-projection}, the vector $\bm{\Gamma}(S,p,\bar{p})$ \eqref{eq:finite-gamma}, and to put them together for the vector $\bm{U}(p)$ \eqref{eq:u-vector}. Once $\bm{U}(p)$ is available, the Poisson-bracket expressions are used once more, to finally evaluate the collisional rate-of-change of the particle $\bm{z}_p$ coordinates via \eqref{eq:collisional-motion}.

The approach we have taken to discretize the collision operator with particles might be somewhat unfamiliar to the readers for the method is deterministic yet involves particles. While it is more common to think of second-order differential operators---also the collision operator discussed here can be written down in the Fokker-Planck form\cite[see Eqs. 4.25 and 4.26]{Differential_formulation:2017}---as generators of diffusion and hence of stochastic processes, inspecting the topic via the metriplectic formulation enables an intuitive interpretation of the collisions as a flow along a {\it compressible} vector field driven by the gradient of the entropy, fully analogous to the {\it incompressible} flow generated by the gradient of the Hamiltonian. This is a trick employed, e.g., in solving diffusion equations with a deterministic particle approach: instead of $\partial_t c=\nabla\cdot (k\nabla c)$, one writes $\partial_tc+\nabla\cdot((-k\nabla\ln c) c)=0$ and interprets $\bm{v}=-k\nabla \ln c$ as a compressible vector field generated by the gradient of the entropy $S=-\ln c$ along which the density $c$ is advected. An excellent discussion and further references of this topic are found in\cite{Carrillo_et_al:2020}.
\section{Summary}
To summarize, we have demonstrated that the modern formulation of collisional electrostatic gyrokinetics exhibits a metriplectic structure and that this structure can be exploited to derive a meaningful and structure-preserving marker-particle approximation of the collision operator. Our findings largely rely on the identity $\delta {\cal H}^{\text{gy}}/\delta F_s=H^{\text{gy}}_s$ which, in case of full-$F$ electromagnetic gyrokinetics, no longer holds, rendering a simple guessing process to derive a collision operator for electromagnetic gyrokinetics difficult. Our new results nevertheless suggest that collisional gyrokinetics, and other dissipative physical models that obey the Laws of Thermodynamics, may be obtained using an as-yet undiscovered metriplectic reduction theory and that theory would likely be useful not only for theoretical considerations but also for numerical implementations. Theory of metriplectic reduction, if it exists, would effectively expand the powerful Lagrangian and Hamiltonian reduction methods into physical systems that display the First and Second Laws of Thermodynamics by enabling one to consistently apply, e.g, the perturbation tools common in Lagrangian reduction while retaining the mathematical stuctures intact also after truncating the perturbative series.


\section*{Data availability statement}
This work presents no data. 

\begin{acknowledgments}
This research was supported by the Academy of Finland grant no. 315278 and by the Los Alamos National Laboratory LDRD program under project number 20180756PRD4. Any subjective views or opinions expressed herein do not necessarily represent the views of the Academy of Finland, Aalto University, or Los Alamos National Laboratory.
\end{acknowledgments}

\bibliography{bibfile.bib}

\end{document}